\documentclass[
    ,final            
  ]
  {aipproc}

\def\power#1{\mbox{$\times10^{#1}\ $}}
\newcommand{\gap}{\mathrel{ \rlap{\raise.5ex\hbox{$>$}}
                    {\lower.5ex\hbox{$\sim$}}  } }
\newcommand{\lap}{\mathrel{ \rlap{\raise.5ex\hbox{$<$}}
	            {\lower.5ex\hbox{$\sim$}}  } }

\newcommand{\fo}{$^{18}$F}
\newcommand{\fn}{$^{19}$F}
\newcommand{\nen}{$^{19}$Ne}

\newcommand{\pa}{$^{18}$F(p,$\alpha)^{15}$O}

\newcommand{\zaa}{{\it Astron. Astrophys.}}

\newcommand{\zapj}{{\it Astrophys. J.}}
\newcommand{\znp}{{\it Nucl.~Phys.}}

\newcommand{\zpr}{{\it Phys.~Rev.}}

\newcommand{\znim}{{\it Nucl.~Inst.~and~Meth.}}

\newcommand{\zmnras}{{\it MNRAS}}

\layoutstyle{6x9}

\begin{document}

\title{A new experiment for the determination of the $^{18}$F(p,$\alpha$)
reaction rate at nova temperatures}

\author{N.~de~S\'er\'eville}{
  address={CSNSM, CNRS/IN2P3/UPS, B\^at.~104, 91405 Orsay Campus, France}
}

\author{A.~Coc}{
  address={CSNSM, CNRS/IN2P3/UPS, B\^at.~104, 91405 Orsay Campus, France}
}

\author{C.~Angulo}{
  address={CRC and FYNU, UCL, Chemin du Cyclotron 2, B-1248 Louvain La Neuve, 
  Belgium}
}

\author{M.~Assun\c{c}\~ao}{
  address={CSNSM, CNRS/IN2P3/UPS, B\^at.~104, 91405 Orsay Campus, France}
}

\author{D.~Beaumel}{
  address={Institut de Physique Nucl\'eaire, CNRS/IN2P3/UPS, 
  91406 Orsay Campus, France}
}

\author{B.~Bouzid}{
  address={USTHB, B.P. 32,  El-Alia, Bab Ezzouar, Algiers, Algeria}
}

\author{S.~Cherubini}{
  address={CRC and FYNU, UCL, Chemin du Cyclotron 2, B-1248 Louvain La Neuve, 
  Belgium}
}

\author{M.~Couder}{
  address={CRC and FYNU, UCL, Chemin du Cyclotron 2, B-1248 Louvain La Neuve, 
  Belgium}
}

\author{P.~Demaret}{
  address={CRC and FYNU, UCL, Chemin du Cyclotron 2, B-1248 Louvain La Neuve, 
  Belgium}
}

\author{F.~de~Oliveira~Santos}{
  address={GANIL, B.P. 5027, 14021 Caen Cedex, France}
}

\author{P.~Figuera}{
  address={Laboratori Nazionali del Sud, INFN, Via S. Sofia, 44 - 95123
Catania, Italy}
}

\author{S.~Fortier}{
  address={Institut de Physique Nucl\'eaire, CNRS/IN2P3/UPS, 
  91406 Orsay Campus, France}
}

\author{M.~Gaelens}{
  address={CRC and FYNU, UCL, Chemin du Cyclotron 2, B-1248 Louvain La Neuve, 
  Belgium}
}

\author{F.~Hammache}{
  address={GSI mbH, Planckstr. 1, D-64291 Darmstadt, Germany}
}

\author{J.~Kiener}{
  address={CSNSM, CNRS/IN2P3/UPS, B\^at.~104, 91405 Orsay Campus, France}
}

\author{D.~Labar}{
  address={Unite de Tomographie Positron, UCL, Chemin du Cyclotron 2, B-1248 Louvain La Neuve, 
  Belgium}
}

\author{A.~Lefebvre}{
  address={CSNSM, CNRS/IN2P3/UPS, B\^at.~104, 91405 Orsay Campus, France}
}

\author{P.~Leleux}{
  address={CRC and FYNU, UCL, Chemin du Cyclotron 2, B-1248 Louvain La Neuve, 
  Belgium}
}

\author{A.~Ninane}{
  address={CRC and FYNU, UCL, Chemin du Cyclotron 2, B-1248 Louvain La Neuve, 
  Belgium}
}

\author{M.~Loiselet}{
  address={CRC and FYNU, UCL, Chemin du Cyclotron 2, B-1248 Louvain La Neuve, 
  Belgium}
}

\author{S.~Ouichaoui}{
  address={USTHB, B.P. 32,  El-Alia, Bab Ezzouar, Algiers, Algeria}
}

\author{G.~Ryckewaert}{
  address={CRC and FYNU, UCL, Chemin du Cyclotron 2, B-1248 Louvain La Neuve,
  Belgium}
}

\author{N.~Smirnova}{
  address={Instituut voor Kern en Stralingsfysika
  Celestijnenlaan 200D, B-3001, Leuven, Belgium}
}

\author{V.~Tatischeff}{
  address={CSNSM, CNRS/IN2P3/UPS, B\^at.~104, 91405 Orsay Campus, France}
}

\author{J.-P.~Thibaud}{
  address={CSNSM, CNRS/IN2P3/UPS, B\^at.~104, 91405 Orsay Campus, France}
}

\begin{abstract}
The \pa\ reaction was recognized as one of the most
important for gamma ray astronomy in novae as it governs the early 511~keV
emission.
However, its rate remains largely uncertain at nova temperatures. A direct
measurement of the cross section over the full range of nova energies is
impossible because of its vanishing value at low energy and 
of the short \fo\ lifetime.
Therefore, in order to better constrain this reaction rate, we have performed an
indirect experiment taking advantage of the availability of a high purity and 
intense radioactive \fo\ beam at the Louvain~La~Neuve RIB facility. 
We present here the first results of the data analysis and discuss the 
consequences.
\end{abstract}

\maketitle


\section{Introduction}

Gamma--ray emission from classical novae is dominated, during the first
hours, by positron annihilation resulting from the beta decay of
radioactive nuclei.
The main contribution comes from the decay of \fo\ (half--life of 110~mn) 
and hence is directly related to \fo\ formation during the outburst. 
(See the astrophysical discussions in references \cite{Gom98,Her99,F00} and by 
Hernanz in these proceedings.)
A good knowledge of the
nuclear reaction rates of production and destruction of \fo\ is required
to calculate the amount of \fo\ synthesized in novae and the resulting 
gamma--ray emission.
The rate (see ref.~\cite{WK82}) relevant for the main mode of \fo\ destruction 
(i.e, through \pa) has been the object of many recent experiments\cite{Gra00,Bar01}
(see also Bardayan in these proceedings and refs. in \cite{F00}).
However, this rate remains poorly known at nova temperatures (lower than
3.5\power{8}~K) due to the scarcity of spectroscopic information for levels
near the proton threshold in the compound nucleus \nen.
This uncertainty is directly related to the unknown proton widths ($\Gamma_p$)
of the first three levels ($E_x$, $J^\pi$ = 6.419~MeV, 3/2$^+$; 
6.437~MeV, 1/2$^-$ and 6.449~MeV, 3/2$^+$).
The tails of the corresponding resonances (at respectively $E_R$ = 8~keV, 26~keV and 38~keV) 
can dominate the astrophysical factor in the relevant energy range\cite{F00}. 
As a consequence of these nuclear uncertainties, the \fo\ production in nova and the 
early gamma--ray emission is uncertain by a factor of 300\cite{F00}.
This supports the need of new experimental studies to improve the reliability
of the predicted annihilation gamma--ray fluxes from novae.

\section{Experiment}

A direct measurement of the relevant resonance strengths is impossible
because they are at least ten orders of magnitude smaller than the weakest
directly measured one (at $E_R=330$~keV;\cite{Gra97} and Bardayan, these proceedings)
due to Coulomb barrier penetrability.
Hence, we used an indirect method aiming at the determination of the one
nucleon spectroscopic factors ($S$) in the analog levels of the mirror nucleus
(\fn) by a neutron transfer reaction: D($^{18}$F,p)$^{19}$F.
(Analog, levels expected to have similar nuclear properties have been
identified in \fn\ and \nen\ spectra\cite{Utk98}.)
From the spectroscopic factors it is possible to calculate the proton widths
through the relation $\Gamma_p=S\times\Gamma_{\mathrm s.p.}$ where
$\Gamma_{\mathrm s.p.}$ is the single particle width readily obtained from a model.
The main reason for the choice of a transfer reaction is the much higher
reaction cross-section as compared to the direct proton capture.
The spectroscopic factors, $S$, are extracted from the angular distribution of the
escaping nucleon via the relation:

\begin{equation}
\label{DWBA}
\left( \frac{d \sigma }{d\Omega} \right)_{exp} = C^2S
\left( \frac{d \sigma }{d\Omega} \right)_{DWBA}
\label{q:dwba}
\end{equation}

Where the $\left({d \sigma }/{d\Omega} \right)_{exp}$ is the experimental
angular distribution of the protons from the D($^{18}$F,p)$^{19}$F reaction while
$\left({d \sigma }/{d\Omega} \right)_{DWBA}$ is the theoretical one
(Distorded Wave Born Approximation) and $C^2$ is a known coefficient.

\begin{figure}[h]
  \includegraphics[width=13cm]{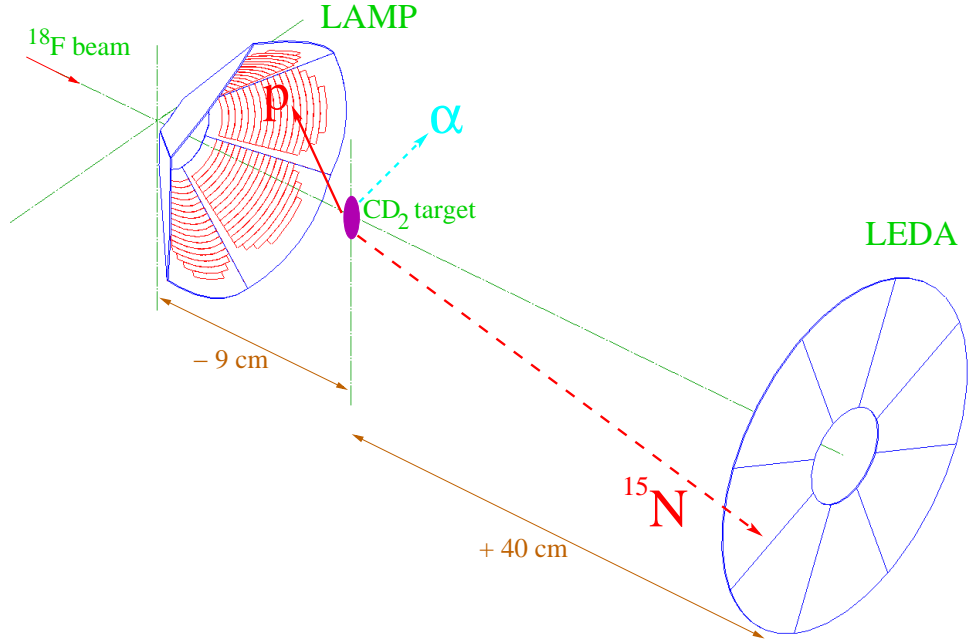}
  \caption{Experimental setup.}
\label{f:setup}
\end{figure}

Since \fo\ is a short lived (110~mn) radioactive isotope, it cannot be used as a
target. It must be first produced, then accelerated and directed to the
deuterium target (inverse kinematics).
We performed the experiment at the {\it Centre de Recherche du Cyclotron} in
Louvain--La--Neuve (Belgium) where such a beam has been developed.
The \fo\ is produced through the $^{18}$O(p,n) reaction, chemically extracted to
form CH$_3^{18}$F molecules, transferred to the cyclotron source\cite{Cog99} and
accelerated to 14~MeV.
The targets are made of deuteriated polypropylene (CD$_2$) of
$\approx$100~$\mu$g/cm$^2$ thickness.
For the energy considered here (1.4~MeV in the center of mass), the deuteron and 
the outgoing proton are both below the Coulomb barrier.
The major advantages is a reduction of the contribution of compound-nucleus 
reactions leading to a better extraction of spectroscopic factors.
The experimental setup is depicted in Figure~\ref{f:setup}. 
It consists of two silicon multistrip detectors composed of sectors with 16 
concentric strips (of 5~mm width) built by the Louvain--La--Neuve and 
Edinburgh collaboration\cite{Dav00}.
They measure the angle (strip number), energy and time of flight (for particle
identification) of the particles.
One, LAMP, is positioned 9~cm upstream from the target; it consists of 6 sectors 
forming a conical shape to optimize angular coverage.
With such a geometry, it covers laboratory angles between 115$^\circ$ and
160$^\circ$ i.e. forward center of mass angles between 12$^\circ$ and 40$^\circ$
providing a good acceptance for protons in the domain of interest for the
differential cross section.
Indeed, the proton angular distribution as measured in LAMP is the
$\left({d \sigma }/{d\Omega} \right)_{exp}$ term in eq.~\ref{q:dwba}.
The other detector, LEDA, is made up of 8 sectors forming a disk
positioned 40~cm downstream from the target and is used for background
reduction and normalization.
The levels of interests are situated high above the alpha emission threshold
(at 4.013~MeV) and their almost exclusive decay mode is through
$^{19}$F$^*\to^{15}$N+$\alpha$.
Hence, to reduce background, we required coincidences between a proton in LAMP and a
$^{15}$N (or $\alpha$ discriminated by time of flight) in LEDA.
Following Monte Carlo simulations, the exact positions of the two detectors have 
been chosen to optimize resolution and acceptance. 
The proton detection efficiency is found to be 24\% and is only slightly reduced 
to 19\% when the coincidence condition is applied.
Rutherford elastic scattering of $^{18}$F on Carbon from the target, detected in LEDA,
provide the (target thickness) $\times$ (beam intensity) normalization.

\section{Results}

\begin{figure}
\includegraphics[width=13cm]{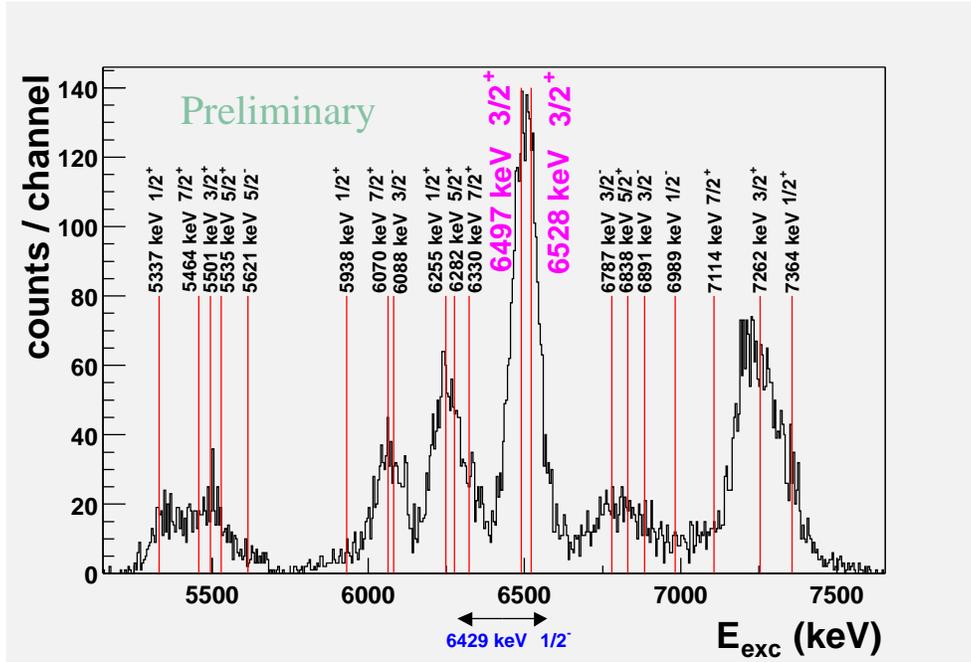}
\caption{Reconstructed \fn\ spectrum (corresponding to 65\% of the total
statistics) showing the two 3/2$^+$ levels of
astrophysical interest around 6.5~MeV of excitation energy.}
\label{f:spect}
\end{figure}

\begin{figure}
\includegraphics[width=13cm]{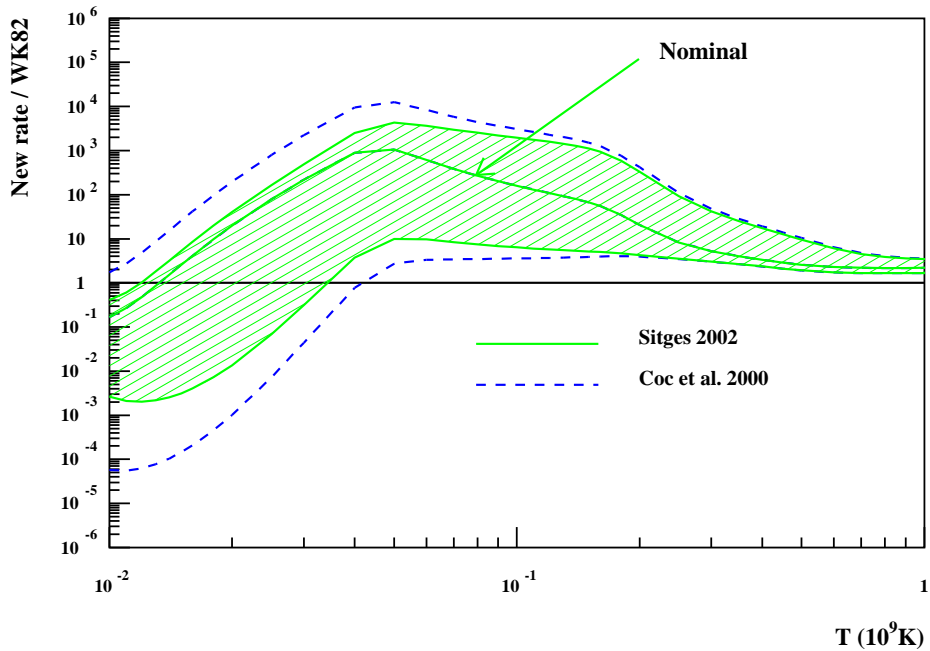}
\caption{Present reduction on rate uncertainties (hatched area) brought by
the experiment compared with previous limits\protect\cite{F00}.
(Ratios are with respect to the Wiescher and Kettner rate\protect\cite{WK82}.)
Note that part of the remaining uncertainty is due to the 1/2$^-$ resonance.
}
\label{f:rate}
\end{figure}

During the 7 days experiment, 15 bunches of $\lap$1~Ci of \fo\ were produced
providing each a mean beam intensity of 5$\times10^6$ particles per second over
a period of $\approx$2~hours.
The beam contamination (by $^{18}$O) was found to be smaller than 10$^{-3}$.
Thanks to the kinematics, at this low energy, only light particles (p and
$\alpha$ from D($^{18}$F,p)$^{19}$F and D($^{18}$F,$\alpha)^{16}$O) can reach LAMP while the
coincidences with LEDA provide a further selection.
The excitation energy of the decaying \fn\ levels can be
kinematically reconstructed from the energies and angles of the detected
protons and the known beam energy.
The corresponding spectrum is represented in Figure~\ref{f:spect} where
vertical lines represent the known position of the \fn\ levels.
The resolution is not sufficient to separate the various levels but the
two 3/2$^+$ levels of interest at 6.497 and 6.528~MeV (the analogs of
the  3/2$^+$ levels in \nen) are well separated from the other groups of levels.
There is no peak corresponding to the 1/2$^-$ level because it is so broad
($\Gamma_T=220$~keV) that it cannot be disentangled from the background.
The angular distribution, $\left({d \sigma }/{d\Omega} \right)_{exp}$, obtained
from the data corresponding to the 6.5~MeV peak, i.e. the 3/2$^+$ levels, is in good
agreement\cite{Ser02} with the theoretical one
$\left({d \sigma }/{d\Omega} \right)_{DWBA}$ (using nuclear potentials
from ref.~\cite{Lop64}) providing evidence that the analysis is reliable
(e.g. negligible compound nucleus contribution and $\ell=0$ transferred
angular momentum).
Since the two 3/2$^+$ levels are not resolved, eq.~\ref{q:dwba} gives the
{\em sum} of the two spectroscopic factors: $S_1+S_2\approx0.2$.
The important consequence of this preliminary value is that the contribution
of these resonances to the rate {\em cannot} be neglected but that the
nominal rate ($S_1=S_2\approx0.1$) used in gamma--ray flux calculations
is not ruled out. However, the extreme case where $S_1\approx0.2$, $S_2=0$
and $S_1=0$, $S_2\approx0.2$ have also to be considered to obtain upper and
lower rate limits.
Figure~\ref{f:rate} shows the present reduction on \pa\ rate uncertainty
brought by this experiment. Hopefully, progress in the data analysis
(energy calibration and normalization) will further reduce this uncertainty but new
experiments are required to obtain a reliable reaction rate for nova gamma--ray 
flux calculations.

\begin{theacknowledgments}
We thank Alan Shotter and his team for allowing us to use the LEDA
and LAMP detectors from the Louvain-La-Neuve and Edinburgh collaboration.
\end{theacknowledgments}

\bibliographystyle{aipproc}   

\begin{thebibliography}{99}


\bibitem{Gom98} G\'omez--Gomar,~J., Hernanz,~M., Jos\'e,~J.
and Isern,~J., \zmnras\ {\bf 296}, 913 (1998). 

\bibitem{Her99} Hernanz,~M., Jos\'e,~J., Coc,~A., G\'omez--Gomar,~J. and
Isern,~J., \zapj\ {\bf 526}, L97 (1999).

\bibitem{F00} Coc,~A., Hernanz,~M., Jos\'e,~J. and Thibaud,~J.P., 
\zaa\ {\bf 357}, 561 (2000). 

\bibitem{WK82}
Wiescher~M. and Kettner,~K.U., \zapj\ {\bf 263}, 891 (1982).

\bibitem{Gra00}
Graulich,~J.-S., Cherubini,~S., Coszach~R., et al.,
\zpr\ {\bf C63}, 011302(R) (2000).

\bibitem{Bar01}
Bardayan,~D.W., Blackmon,~J.C., Bradfield--Smith,~W. et al.,
\zpr\ {\bf C63}, 065802 (2001).

\bibitem{Gra97}
Graulich,~J.-S., Binon,~F., Bradfield--Smith,~W. et al.,
\znp\ {\bf A626}, 751 (1997).

\bibitem{Utk98}
Utku,~S., Ross,~J.G., Bateman,~N.P.T. et al.,
\zpr\ {\bf C57}, 2731 and {\bf C58}, 1354 (1998).

\bibitem{Cog99} Cogneau,~M., Decrock,~P., Gaelens,~M., Labar,~D. et al., \znim\ 
{\bf A420}, 489 (1999). 

\bibitem{Dav00} Davinson,~T., Bradfield--Smith,~W. Cherubini,~S. et al., \znim\
{\bf A454}, 350 (2000).

\bibitem{Ser02} de~S\'er\'eville,~N. et al., in preparation and in
``7$^{\mathrm th}$ Internal Symposium on Nuclei in the Cosmos'', proceedings to appear
in {\it Nucl.~Phys.~A}.

\bibitem{Lop64} de~L\'opez,~M.E.O., Rickards,~J. and Mazari,~M.,
\znp\ {\bf 51}, 321 (1964).

\end{thebibliography}

\end{document}